# Present value of the future consumer goods multiplier

**Ihor Kendiukhov**
Faculty of Economics and Business Administration, Humboldt University of Berlin
Unter den Linden 6, 10099 Berlin
Germany
e-mail: kenduhov.ig@gmail.com

# Present value of the future consumer goods multiplier

**ABSTRACT:** In this paper, we derive a formula for the present value of future consumer goods multiplier based on the assumption that a constant share of investment in the production of consumer goods is expected. The present value appears to be an infinite geometric sequence. Moreover, we investigate how the notion of the multiplier can help us in macroeconomic analysis of capital and investment dynamics and in understanding some general principles of capital market equilibrium. Using the concept of this multiplier, we build a macroeconomic model of capital market dynamics which is consistent with the implications of classical models and with the market equilibrium condition but gives additional quantitative and qualitative predictions regarding the dynamics of shares of investment into the production of consumer goods and the production of means of production. The investment volume is modeled as a function of the multiplier: investments adjust when the value of the multiplier fluctuates around its equilibrium value of one. In addition, we suggest possible connections between the investment volume and the multiplier value in the form of differential equations. We also present the formula for the rate of growth of the multiplier. Independently of the implications of capital market dynamics models, the formula for the multiplier itself can be applied for the evaluation of the present value of capital or the estimation of the macroeconomic impact of changes in investment volumes. Our findings show that both the exponential and hyperbolic discounting in combination with empirical evidence available lead to the value of the multiplier that is close to one.

## 1. INTRODUCTION

The present value method of asset pricing is one of the most generally accepted and widely used in finance and macroeconomics (see e.g. Berk and Van Binsbergen, 2016; or Pohl, Schmedders, Wilms, 2018). For example, Gallo (2014) provides a comprehensive review of the state of the subject and lists common assumptions of the present value calculation process which we also use. Another review is made by Fernandez (2019). Generality and applicability of the net present value criterion for similar problems are shown by Graham and Harvey (2001) or Brounen et al. (2004). Although our analysis is macroeconomic in its essence, it is crucial to understand the role the net present value (NPV) plays on a level of a single firm, from the standpoint of corporate finance. This is investigated by Pasqual, Padilla, and Jadotte (2013). In addition, Bierman and Smidt (2012) studied capital budgeting from various points of view, contributing to the understanding of the optimality criteria for investment decision. The multiplier we derive can be applied both in the world of efficient markets and outside of it. Actually, the multiplier itself can be applied for testing the efficient market hypothesis. However, the efficient market hypothesis is needed to make some further conclusions out of the multiplier formula and in order to investigate equilibrium values of multiplier and its equilibrium behavior. An overview of the efficient market hypothesis relevant to this task is made by Degutis and Novickytė (2014), as well as by Gabriela ğiĠan, A. (2015).

In this paper, we present a new formula for the estimation of the present value of a unit of capital on the macroeconomic level and make some implications regarding its parameters and consistency with standard macroeconomics theory. The formula is derived as a sum of eternal cash flows generated from the initial investment, assuming that one constant share of investment flows into the production of means of production, whereas another constant share of investment flows into the production of consumer goods. Technically, it can be described as the present value of some multiplier M. This multiplier, in turn, calculates the total future value of all consumer goods that will be created by the initial investment of one unit of capital.

We derive the value of future consumer goods multiplier based on the analysis of the sequence of consumer goods produced in the future by one unit of investment and investigate parameters it depends on. Then, we investigate how the notion of the multiplier can help us in macroeconomic analysis of capital and investment dynamics and in understanding some general principles of capital market equilibrium. Namely, it is shown that analysis of the multiplier function leads to theoretically relevant results and that changes in capital stock can be described as a result of the adjustment of the multiplier to the value of one. Being consistent with modern macroeconomic theory, the model brings new highlights to the problem of capital accumulation. The multiplier is also investigated in the connection with the intertemporal choice theory. The main research question of this paper is the following one: how can we measure the present value of the average capital unit invested?

Then, the dynamical behavior of the multiplier formula is analyzed in order to make a prediction regarding the investment and capital market dynamics. It is shown that the notion of the present value of future consumer goods multiplier can be helpful for understanding the impact of discount rate and marginal productivity of capital on capital markets and investment volumes.

This paper is structured as follows: Section 2 provides a comprehensive literature review. Section 3 outlines the research methodology describing the materials and methods, as well as the model itself. Section 4 provides the main results and outcomes. Section 5 summarizes our main findings with a discussion. Finally, section 6 lists the main findings and implications.

## 2. LITERATURE REVIEW

It is a well-known fact that the issue of present value is deeply investigated in modern economic theory. The difference between efficient and inefficient capital markets for the multiplier theory is very important, since in the case of efficient capital markets we prove that the multiplier is consistent with market equilibrium and fluctuates at the value of one. In the case of inefficient capital markets, the value of the multiplier may not adjust to one, and the dynamic models will be less applicable. For the study of the discussion on market efficiency, we refer to Akbas, Armstrong, Sorescu, Subrahmanyam (2016), to LeRoy and Lansing (2016), as well as to Williams, Dobelman, (2017). Various empirical case studies deliver important insights as well (Boya, 2017; Kumar, Raman, 2016; or Ma, 2017).

As we refer to the concept of marginal productivity of capital in the study, it is important to define it properly (Fuller, 2013). Marginal productivity of capital theory is tested empirically by Biewen and Weiser (2014). The limitations of the theory are studied by Moseley (2012).

Intertemporal choice is another area that relates directly to the problem we analyze. Holistically the problem of intertemporal choice was studied by Loewenstein, Read, Baumeister (2003), Chabris, Laibson and Schuldt (2010). They enlighten economic, psychological and neurobiological approaches to the question of what is extremely helpful in applying different models of time discounting and understanding psychological of the phenomenon itself. Psychological aspects of discounting are studied by Matta, Gonçalves and Bizarro (2012).

The works dedicated to the macroeconomic theory of investment and capital are of huge importance for our problem, starting from such classic authors as Fisher (1906) and ending with Garisson (2001), Fuster, Hebert, Laibson (2012). The recursive macroeconomic theory brings similar ideas of capital dynamics (Ljungqvist, Sargent, 2018).

Of course, the very concept of multiplier dates back to Keynes (1936). We build the present value of capital multiplier following the common notion of the macroeconomic multiplier, which is developed in the works of Gnos and Rochon (2008), Ono (2011), Cogan, Cwik, Taylor, and Wieland (2010), Westerhoff (2006), Robinson (2006), Currie (1983), Smithies (1948).

Once derived, the multiplier is applied to the analysis of macroeconomic problems of capital accumulation, investment and interest rates dynamics, and financial markets. Hamouda (2011) delivers a survey of modern Keynesian approaches to the problems of investment on the macro level, which is important for the analysis of the connection of multiplier with interest rates dynamics. However, the multiplier itself is constructed more in the spirit of neoclassical models, where the concept of marginal productivity and long-term equilibrium plays an important role. The equilibrium concepts we use are based on the results of Caselli and Feyrer (2007).

Although the applicability and efficiency of the notion of the Keynesian multiplier are in general questionable (Carpenter and Demiralp, 2012; Castelnuovo and Lim, 2019), it does not affect the applicability of the multiplier we present, since it is not a demand-side multiplier and all demand-side critique cannot be applied. The research in the field, however, focuses more on fiscal, rather than investment, multipliers (Spilimbergo, Schindler, Symansky, 2009, Dupor and Guerrero, 2017). Whereas we model the investment demand, the derivation of the multiplier formula is conducted in the supply-side spirit (Vollet, Aubert, Frère, Lépicier, Truchet, 2018).

Our differential equations models can be analyzed and compared in the context of modern theories of capital dynamics. Although we do not study explicitly the implications of the model for monetary and fiscal policy, such analysis can be conducted based on the frameworks described by Laopodis (2013), Borio and Zhu (2012), Bekaert, Hoerova and Duca (2013), Arrow and Kruz (2013), Bhattarai and Trzeciakiewicz (2017). Models constructed based on the dynamic stochastic general equilibrium (Khramov, 2012; Herbst and Schorfheide, 2015) may have a similar form. The boundaries of applicability of the capital dynamics model presented in the paper can be estimated by the study of the model under the conditions of inefficient capital markets. Such capital market imperfections are listed, for example, by Rotheim (2013).

3. **METHODOLOGY AND MODEL**

The present value of future consumer goods multiplier (later denoted as *M*) is constructed technically similar to common Keynesian multipliers (calculating the value of infinite geometric sequence), but economically its value is created on supply, not demand, side. In this sense, it is rather neoclassical, than a Keynesian, multiplier.

Let us define variables: *K* – initial capital investment; *p* – marginal productivity of capital; *n* - the rate of depreciation; *a=1-n*; *c* – the share of investment in the production of consumer goods; *i* – the share of investment in the production of means of production; *R* – discount rate; *r=1/(1+R)* (discount factor).

The amount of capital K=1 is invested in year 0. It is divided into 2 shares – one goes into consumer goods production and another in production of means of production. Thus, we can build the following structure of goods produced in each year:

$$\begin{cases} t_1: (cp)+ip \\ t_2: (cpa+cip^2)+ipa+i^2p^2 \\ t_3: (cpa^2+cip^2a+cp^2ai+ci^2p^3)+ipa^2+i^2p^2a+i^2p^2a+i^3p^3 \\ \quad \dots \\ t_n: (cpa^n+\dots+ci^np^{n+1})+ipa^n+\dots+i^{n+1}p^{n+1} \\ \quad \dots \end{cases} \qquad (1)$$

However, one should note that we are interested only in consumer goods. They are highlighted by brackets. We can consider them separately. Hence, the total amount of consumer goods produced in the future by 1 unit of capital investment is following:

$$\begin{cases} cp \\ + \\ cpa+cip^2 \\ + \\ cpa^2+cip^2a+cp^2ai+ci^2p^3 \\ + \\ cpa^3+cip^2a^2+cp^2a^2i+ci^2p^2a+cp^2a^2i+ci^2p^3a+cp^3ai^2+ci^3p^4 \\ + \\ \dots \\ + \\ cpa^n+\dots+ci^np^{n+1} \\ + \\ \dots \end{cases} \qquad (2)$$

We can observe that the sequence of lines builds geometric sequence:

$$S_{n+1} = S_n(a + ip) \qquad (3)$$

where $S_n$ is line n in term (2).

If $a + ip < 1$, then the value of the term (2) is the sum of infinitely decreasing geometric sequence:

$$M = \frac{cp}{1-a-ip} \tag{4}$$

Let us call M future consumer goods multiplier. The present value of future consumer goods multiplier can be thus presented as following:

$$M_r = \frac{cpr}{1-r(a+ip)} = \frac{cpr}{1-ra-rp+rpc} \tag{5}$$

It is natural now to attempt to find extreme points of the function $M_r(c)$:

$$\frac{\partial M_r}{\partial c} = \frac{rp - r^2p^2 - ar^2p}{(1-pr-ar+rpc)} \tag{6}$$

Term (6) is never equal to 0.

As we know, expression $M_r$ is a hyperbolic function with respect to c. Thus, since term $pr$ is always positive, the local optimum of the function $M_r(c)$ on the interval $c \in [0; 1]$ is:

$$c^* = \begin{cases} 0, (1-ra-rp) < 0 \\ (0;1), (1-ra-rp) = 0 \\ 1, (1-ra-rp) > 0 \end{cases} \tag{7}$$

It is definitely true that in the real economy $c^* \epsilon (0; 1)$. Then:

$$p = R + n \tag{8}$$

The empirical validity of this expression can be tested. Yet this expression is just another form of famous result MP=MC, which shows the theoretical integrity of the calculations made.

If expression (8) is true, then formula (5) for multiplier $M_r$ is consistent with market equilibrium condition for any possible values of parameters. Indeed, we know that under the law of one price:

$$PV(K) \equiv K * M_r = K \frac{cpr}{1-r(a+ip)} = K \tag{9}$$

Then:

$$M_r = \frac{cpr}{1-r(a+ip)} = 1 \tag{10}$$

If the capital market is in equilibrium, $M_r$ should be equal to 1.

If $p = R + n$, then:

$$M_r = \frac{cpr}{1-r(a+ip)} = \frac{c(R+n)\frac{1}{1+R}}{1-\frac{1}{1+R}(1-n+(1-c)(R+n))} = \frac{\frac{cR}{1+R}+\frac{cn}{1+R}}{\frac{1+R}{1+R}-\frac{1-n}{1+R}-\frac{(1-c)(R+n)}{1+R}} = \frac{cR+cn}{1+R-1+n-R-n+cR+cn} =$$

$$\frac{cR+cn}{cR+cn} = 1 \tag{11}$$

It turns out that (5) and (8) are equivalent in terms of market equilibrium definition. Condition (10) follows directly from the equilibrium (8).
We can rewrite expression (7) in the following form:

$$c^* = \begin{cases} 0, \ p > R + n \\ (0; 1), \ p = R + n \\ 1, p < R + n \end{cases} \tag{12}$$

It means that it is optimal to invest in the production of consumer goods if $p < R + n$, and it is optimal to invest in the production of means of production if $p > R + n$.
Term (12) also tends to be justified if we analyze the issue from the standpoint of consumer choice problem:

$$maxU = C + M_r K \ s.t. W = C + K \tag{13}$$

where C is today consumption, W is today wealth.

It is obvious that we get a corner solution. If $M_r > 1$, then K=W, if $M_r < 1$, then K=0. And only if $M_r = 1$, $K \in (0; W)$. Only the third case is empirically relevant. Let us assume that p adjusts, so that (8) is true:

$$\frac{dK}{dt} = p(K,t) - R - n \tag{14}$$

This differential equation describes the net capital change. Then, gross capital change (investment expenditure) is:

$$I = \frac{dK_g}{dt} = \frac{dK}{dt} + n = p(K,t) - R \tag{15}$$

In this case, we will observe investment volume fluctuations around the value $p^{-1}(R + n)$, $p^{-1}$ is the reverse function of p. Let us use the Cobb-Douglas function:

$$Y = AL^a K^b \tag{16}$$

Then, for A, L=const, the solution of equation (14) is:

$$K = (x \to \frac{x \, _2F_1\left(1, \frac{1}{b-1}, 1+\frac{1}{b-1}, \frac{AbK^{b-1}L^a}{n+R}\right)}{n+R})^{-1} (c_1 - t) \tag{17}$$

where F is the hypergeometric function.

The solution of equations (15) is corresponding:

$$K = (x \to \frac{x \, _2F_1\left(1, \frac{1}{b-1}, 1+\frac{1}{b-1}, \frac{AbK^{b-1}L^a}{R}\right)}{R})^{-1} (c_1 - t) \tag{18}$$

But for the long-term analysis, A and L should not be treated as constants. Then, for the solution of the equations, numerical methods can be applied.

Let us consider the case of hyperbolic discounting (Grüne-Yanoff, Till, 2015, Hampton, Venkatraman, Olson, 2017). Then, the present value of future consumer goods produced in period n is:

$$S_n = \frac{cp(a+ip)^{n-1}}{1+kn} \tag{19}$$

For $(a + ip) < 1$:

$$M_r = \frac{cp}{k} \phi\left(a + ip, 1, 1 + \frac{1}{k}\right) = \frac{cp}{k} \phi\left(a + p - pc, 1, 1 + \frac{1}{k}\right) \tag{20}$$

where $\phi$ is Lerch zeta function (Apostol, 2010). The optimum then is the following:

$$\frac{\partial M_r}{\partial c} = \frac{p\left(-\frac{1}{k}-1\right)\phi\left(a+p-cp, 1, 1+\frac{1}{k}\right) + \frac{1}{1-a-p+pc}}{a+p-pc} = 0 \tag{21}$$

We can also consider the case of continuous-time:

$$M_r = \int_0^\infty \frac{cp(a+ip)^{n-1}}{1+kn} dn = \lim_{n\to\infty} \frac{cp(a+ip)^{-\frac{k+1}{k}} Ei(\frac{(kn+1)\ln(a+ip)}{k})}{k} - \frac{cp(a+ip)^{-\frac{k+1}{k}} Ei(\frac{\ln(a+ip)}{k})}{k} = \infty \tag{22}$$

Obviously, the case of continuous time is irrelevant as it leads to meaningless results. If some permanent growth of marginal productivity of capital is expected (for example, it can be due to technological progress), both in the cases of exponential and hyperbolic discounting the value of the multiplier approaches infinity. But equation (8) should be true, which leads to the compensation of the effect of the growth of p.
Nevertheless, the discount rate can adjust to the new level of marginal productivity of capital only gradually (as soon as the market interest rate does). It means that the growth of p can lead to the situation when $M_r$ is always greater than 1. It would mean the flow of income from consumption to production:

$$\begin{cases} \frac{dR}{dt} = p - R - n \\ M_r = \frac{cpr}{1-ra-rp+rpc} \end{cases} \tag{23}$$

Is it possible for $M_r$ to reach infinity? If the growth of p is exponential, then yes, if condition (24) is true:

$$R = \beta e^{-t} - \frac{n}{g*lnp+1} - \frac{g*n*lnp}{g*lnp+1} + \frac{p^{gt}}{g*\ln p+1} \tag{24}$$

where $\beta$ is some constant.

$$\lim_{t\to\infty}(p - R) = \lim_{t\to\infty}\left(p^t - \beta e^{-t} + \frac{n}{g*lnp+1} + \frac{g*n*lnp}{g*lnp+1} - \frac{p^{gt}}{g*\ln p+1}\right) = \infty \tag{25}$$

$$\lim_{t\to\infty} M_r = \lim_{\substack{p\to\infty \\ R\to\infty}} M_r = \infty \tag{26}$$

Yet the question, what does it mean, to have an "infinite" value, lies beyond the scope of the paper.
Formula (12) gives strange prediction that in the case of the infinite value of $M_r$ share of investment in means of production would be 1. But why would people refuse from a higher value in the near future, even if the total future value is "infinite"?

We observed now 2 options: either marginal productivity of capital or discount rate adjusts to hold the equilibrium. But what is more probable, they do it simultaneously. Obviously, the most simple and evident way of that is:

$$\begin{cases} \frac{dR}{dt} = p - R - n \\ \frac{dp}{dt} = R + n - p \end{cases} \quad (27)$$

Nevertheless, it leads to the result that $p$ always equals $R+n$ and there are no changes in capital, which is obviously not true. We may consider more complicated cases. Firstly, we should note that investment depends on the value of the future goods value multiplier (when the multiplier is greater than 1, it is reasonable to invest):

$$\frac{dK}{dt} = C(M_r - 1) \quad (28)$$

where $C$ is some parameter.

We can also consider the change of the multiplier in a short time interval:

$$\Delta M_r = \frac{\partial M_r}{\partial p} \frac{\partial p}{\partial t} \Delta t + \frac{\partial M_r}{\partial R} \frac{\partial R}{\partial t} \Delta t \quad (29)$$

where:

$$\frac{\partial M_r}{\partial p} = \frac{r(1 - a\frac{1}{1+R})}{(1 - a\frac{1}{1+R} - p\frac{1}{1+R} + \frac{1}{1+R})^2} \quad (30)$$

$$\frac{\partial M_r}{\partial R} = -\frac{cp}{(1 - a - p + r + cp)^2} \quad (31)$$

We know that in the equilibrium:

$$\Delta M_r = 0 \quad (32)$$

hence:

$$\frac{\partial M_r}{\partial p} \frac{\partial p}{\partial t} \Delta t = -\frac{\partial M_r}{\partial R} \frac{\partial R}{\partial t} \Delta t \quad (33)$$

The formula (33) is the fundamental prediction of the model. But $c$ can also be a variable. Then:

$$\Delta M_r = \frac{\partial M_r}{\partial p}\frac{\partial p}{\partial t}\Delta t + \frac{\partial M_r}{\partial R}\frac{\partial R}{\partial t}\Delta t + \frac{\partial M_r}{\partial c}\frac{\partial c}{\partial p}\frac{\partial p}{\partial t}\Delta t + \frac{\partial M_r}{\partial c}\frac{\partial c}{\partial R}\frac{\partial R}{\partial t}\Delta t = \frac{\partial M_r}{\partial p}\frac{\partial p}{\partial t}\Delta t + \frac{\partial M_r}{\partial R}\frac{\partial R}{\partial t}\Delta t + \frac{\partial M_r}{\partial c}\left(\frac{\partial c}{\partial p}\frac{\partial p}{\partial t} + \frac{\partial c}{\partial R}\frac{\partial R}{\partial t}\right)\Delta t \quad (34)$$

However, we should note that when condition (8) holds, $\frac{\partial M_r}{\partial c}\left(\frac{\partial c}{\partial p}\frac{\partial p}{\partial t} + \frac{\partial c}{\partial R}\frac{\partial R}{\partial t}\right)\Delta t$ should be equal to zero. Hence, in competitive markets, formula (34) can be reduced to formula (29).

In the context of the present value of future consumer goods multiplier, in general, various models of investment dynamics can be proposed:

$$\begin{cases} \frac{dK}{dt} = M_r(K,R) - 1 \\ \frac{dR}{dt} = p - R - n \end{cases} \quad (35)$$

or

$$\begin{cases} \frac{dK}{dt} = M_r(K,R) - 1 \\ \frac{dR}{dt} = K - K^* \\ M_r(K^*, R) = 1 \end{cases} \quad (36)$$

or

$$\begin{cases} \frac{dK}{dt} = M_r(K,R) - 1 \\ \frac{dp}{dt} = R + n - p \end{cases} \quad (37)$$

or

$$\begin{cases} \frac{dK}{dt} = M_r(K,R) - 1 \\ R = f(K) \end{cases} \quad (38)$$

## 4. RESULTS AND OUTCOMES

Quantitative estimations of the derived functions based on macroeconomic data should be made. Moreover, the predictions of the model appear to be the following:

1. Investment grows as $M_r$ grows.
2. Investment grows as the discount rate falls.

3. Investment grows as marginal productivity of capital grows.
4. Formula (34) is true.
5. The share of investment in the production of consumer goods falls when $p > R + n$.
6. The share of investment in the production of consumer goods grows when $p < R + n$.
7. The value of the multiplier is close to 1.
8. Ceteris paribus, the share of investment in the production of consumer goods falls when marginal productivity of capital grows or when interest rate falls, and vice versa.
9. Ceteris paribus, the share of investment in the production of means of production grows when marginal productivity of capital grows or when interest rate falls, and vice versa.
10. When the discount rate falls, the rate of growth of investment in the production of means of production should be higher than the rate of growth of investment in the production of consumer goods (actually the later may even be negative).
11. When the discount rate grows, the rate of growth of investment in the production of means of production should be lower than the rate of growth of investment in the production of consumer goods (and the rate of growth of investment in the production of means of production should be negative).

In general, our model is consistent with the present macroeconomic models of capital dynamics (Rode (2012), Blanchard, Fischer (1989), Meng, Yip, (2004), Caselli, Feyrer, (2007), Tan, Tang, (2016), Lin, Wang, Wang, Yang (2018)), namely the predictions 2, 3, and 7 are consistent. The model, however, gives predictions additionally to what is available in the literature. In this sense, the multiplier model can be distinguished from standard models of investment dynamics by additional predictions it makes. The neoclassic capital market model is thus a special case of the multiplier model. Tables 1 and 2 that follow list the differences in the models. In addition, they show the additional predictions of the multiplier model.

**Table 1.** Comparison of the predictions of standard models with the predictions of the multiplier model, if discount rate R falls

|  | Investment | Share of investment in c | Share of investment in i | Rate of growth of investment in c | Rate of growth of investment in i |
|---|---|---|---|---|---|
| Standard | + | NA | NA | NA | NA |
| Models | | | | | |
| Multiplier model | + | - | + | NA | + |

Notation: "i" – means of production, "c" – consumer goods, "NA" – not available (the model cannot give an unequivocal prediction), "+" – the impact is positive, "-" – the impact is negative.
**Source:** Own results

**Table 2.** Comparison of the predictions of standard models with the predictions of the multiplier model, if discount rate R grows.

| | Investment | Share of investment in c | Share of investment in i | Rate of growth of investment in c | Rate of growth of investment in i |
|---|---|---|---|---|---|
| Standard Models | - | NA | NA | NA | NA |
| Multiplier model | - | + | - | NA | - |

Notation: "i" – means of production, "c" – consumer goods, "NA" – not available (the model cannot give an unequivocal prediction), "+" – the impact is positive, "-" – the impact is negative.
**Source:** Own results

There is ambiguity regarding the rate of growth of investment in the production of consumer goods depending on the discount rate. It is due to the assumption that the share of investment in the production of consumer goods growth when investment falls, and vice versa. This ambiguity can be removed by analyzing the dynamics of investment in the production $\Delta K_c$ of consumer goods with respect to R:

$$\Delta K_c = c(R) * \Delta K(R) \tag{39}$$

$$\Delta K_c = \frac{\partial c}{\partial R}\frac{\partial R}{\partial t}\Delta t + \frac{\partial K}{\partial R}\frac{\partial R}{\partial t}\Delta t \tag{40}$$

We know that $\frac{\partial c}{\partial R} > 0$ and $\frac{\partial K}{\partial R} < 0$. Then:

$$\begin{cases} \Delta K_c \geq 0, \ \frac{\partial c}{\partial R} > -\frac{\partial K}{\partial R} \\ \Delta K_c < 0, \ \frac{\partial c}{\partial R} < -\frac{\partial K}{\partial R} \end{cases} \tag{41}$$

Hence, the impact of *R* on the investment in the production of consumer goods depends on the speed of adjustment of c to R relative to the speed of adjustment of *K* to *R*.

## 5. DISCUSSION OF RESULTS

The concept of the multiplier developed in the paper can be applied for the evaluation of the present value under various market regimes. The derived models of capital dynamics, however, work only in the case of efficient capital markets.
Although it is difficult to test some empirical predictions of the model, since the model is consistent with capital market equilibrium (in terms of the theory of marginal productivity of capital), the evidence supporting capital market efficiency is also supporting the multiplier capital dynamics model.
The model can be falsified by testing its additional predictions, i.e. the predictions it gives above the predictions of standard models of investment dynamics.
Although some assumptions of the multiplier model appear to be very simplified (such as constant depreciation rate), given the market if efficient, the changes in these parameters do not affect the results.
Whereas the formula for the multiplier itself works just by definition in the case of market efficiency, the consequent models of investment dynamics may be or be not empirically relevant.
Of course, the model works until the concept of discounted cash flows works. It means that if the value of the discounted future cash flows for some reason diverges, the model is not relevant. The results in formulas 29, 33 and 34 can be considered as the main predictions of the model (beyond its predictions for the impact of the change in the discount rate on the shares and the growth rates of the investment in the production of consumer goods and means of productions respectively). They should be considered, however, as a way to represent the results of standard macroeconomic theories (which are based on the concept of marginal productivity) of capital market dynamics.
Different marginal productivity of capital models can be built to test the consistency of the predictions of our model.
In the case of market inefficiency, the models should be extended to take into account the fact that the marginal productivity of capital condition may not hold.

## 6. CONCLUSIONS AND IMPLICATIONS

Overall, it appears that further research may focus on empirical tests of the proposed models; extending the formula for the multiplier with respect to more complex assumptions; simulation of macroeconomic shocks in order to investigate the fitness of the multiplier model for the description of the investment dynamics and capital markets; study of the capital dynamics model in the case of the inefficient capital market.

It is clear that economic agents do not expect in general the share of investment in the production of consumer goods, as well rate of depreciation, to be constant, but when condition (8) holds (and it should hold in competitive markets), the value of the multiplier does not depend on this share. The changing rate of depreciation can also be integrated in the model, as it was done with the marginal productivity of capital (see formulas 34-38).

Despite its assumptions` simplicity, the multiplier formula seems to work whenever the markets are efficient. The model can be adjusted to various market inefficiencies. For example, it is reasonable to ask what happens with the value of the multiplier if condition (8) does not hold, and how can we justify that it is still $c \in (0; 1)$ in this case. In the paper, predictions of the model are explicitly listed, so the authors hope that they will be tested empirically. Another topic for empirical study is testing of various investment dynamics models (formulas 35-39 and similar) and selection of the most valid among them.

The implications for economic policy may also be studied. Namely, how monetary policy may change if we know that interest rate changes affect investment in different sectors differently. It is also interesting to know who crowding out effect would affect investment in the production of consumer goods or the production of means of production.

The multiplier presented in the paper may be seen as the first approximation of the reality in terms of the present value of the investment. The model may become more precise if more dynamical parameters are added. However, if the market is efficient, the multiplier value is robust to the changes of parameters.

Our analysis of the value of future consumer goods multiplier gives the result that corresponds with the basic implications of economic theory. Dynamics of capital and investment can be described as fluctuations that adjust the value of the multiplier to the value of one. The value of the multiplier can theoretically be infinite and is possible if some prerequisites are satisfied. Both exponential and hyperbolic discounting in combination with empirical evidence available lead to the value of the multiplier that is close to one.

The multiplier model presented in the paper is consistent with common macroeconomic models of capital market equilibrium and investment dynamics and expands them in the sense that it gives additional specific predictions on the issues where standard models fail to make unequivocal statements. Such predictions relate mostly to the dynamics of investment with respect to changes in the discount rate.

Different models of investment dynamics based on the adjustment of the multiplier to the value of one are possible. They should be tested empirically. Independently of the implications of capital market dynamics models, the formula for the multiplier itself can be applied for the evaluation of the present value of capital or the estimation of the macroeconomic impact of changes in investment volumes.